\documentclass[aps,prc,preprint,superscriptaddress]{revtex4-1}
\usepackage{bm}
\usepackage{amsmath}
\usepackage{epstopdf}
\usepackage{graphicx}
\usepackage{float}
\usepackage{color}

\begin{document}

%\preprint{}

\title{Ternary quasifission in collisions of actinide nuclei}


\author{D. D. Zhang}
\affiliation{State Key Laboratory of Nuclear Physics and Technology, School of Physics, Peking University, Beijing 100871, China}

\author{B. Li}
\affiliation{State Key Laboratory of Nuclear Physics and Technology, School of Physics, Peking University, Beijing 100871, China}

\author{D. Vretenar}
\email{vretenar@phy.hr}
\affiliation{Physics Department, Faculty of Science, University of Zagreb, 10000 Zagreb, Croatia}
\affiliation{State Key Laboratory of Nuclear Physics and Technology, School of Physics, Peking University, Beijing 100871, China}

\author{T. Nik\v si\' c}
\affiliation{Physics Department, Faculty of Science, University of Zagreb, 10000 Zagreb, Croatia}
\affiliation{State Key Laboratory of Nuclear Physics and Technology, School of Physics, Peking University, Beijing 100871, China}

\author{Z. X. Ren}
\affiliation{Institut f\"ur Kernphysik, Institute for Advanced Simulation and J\"ulich Center for Hadron Physics, Forschungszentrum J\"ulich, D-52425 J\"ulich, Germany}
\affiliation{Helmholtz-Institut f\"ur Strahlen- und Kernphysik and Bethe Center for Theoretical Physics, Universit\"at Bonn, D-53115 Bonn, Germany}
\affiliation{State Key Laboratory of Nuclear Physics and Technology, School of Physics, Peking University, Beijing 100871, China}

\author{P. W. Zhao}
\email{pwzhao@pku.edu.cn}
\affiliation{State Key Laboratory of Nuclear Physics and Technology, School of Physics, Peking University, Beijing 100871, China}

\author{J. Meng}
\email{mengj@pku.edu.cn}
\affiliation{State Key Laboratory of Nuclear Physics and Technology, School of Physics, Peking University, Beijing 100871, China}


\date{\today}

\begin{abstract}
The microscopic framework of time-dependent covariant density functional theory is applied to a systematic study of ternary quasifission in collisions of pairs of $^{238}$U nuclei. It is shown that the inclusion of octupole degree of freedom in the case of head-to-head collisions, extends the energy window in which ternary quasifission occurs, and greatly enhances the number of nucleons contained in a middle fragment. Dynamical pairing correlations, included here in the time-dependent BCS approximation, prevent the occurrence of ternary quasifission in head-to-head collisions, and have an effect on the location of the energy window in which a middle fragment is formed in tail-to-tail collisions. In the latter case, as well as for tail-to-side collisions, the formation of very heavy neutron-rich systems in certain energy intervals is predicted. 
\end{abstract}

%\keywords{}

\maketitle
\section{Introduction}\label{sec1}

Over the last decades, significant progress has been achieved in the synthesis of neutron-rich atomic nuclei in fission, fragmentation, fusion, and multinucleon transfer reactions~\cite{ThoennessenRPP2013}. Fission and fragmentation are extensively employed for the production of neutron-rich nuclei below uranium. When it comes to heavy nuclei, fusion emerges as the predominant method~\cite{HofmannRMP2000,OganessianPRC2004}. However, due to the curvature of the stability line, fusion reactions with stable projectiles tend to produce neutron-deficient nuclei. In order to synthesize heavy  neutron-rich nuclei and eventually reach the island of stability, it is crucial to explore alternative methods. Recently, multinucleon transfer reactions have attracted considerable interest as a promising approach for the production of new neutron-rich nuclei~\cite{AyikPRC2017,ZagrebaevJPG2007,ZhaoPRC2015,ZhaoPRC2013,GuoPRC2019,JiangPRC2020,WatanabePRL2015,ZagrebaevPRC2013,SekizawaPRC2013,SimenelPRL2010}.

In particular, the low-energy multinucleon transfer process between actinide nuclei, such as two $^{238}$U nuclei, presents a possible pathway for producing neutron-rich actinide and transactinide isotopes. Therefore, the collision $^{238}$U + $^{238}$U has been extensively investigated, both experimentally \cite{GolabekIJMPE2008,KratzPRC2013} and theoretically \cite{TianPRC2008,GolabekPRL2009,ZhaoPRC2009,ZhaoPRC2015,ZhaoPRC2013,ZhaoPRC2016}. 
Using a model based on coupled Langevin-type equations~\cite{ZagrebaevJPG2007}, Zagrebaev et al. predicted that the presence of shell effects in the $^{238}$U + $^{238}$U collision facilitates the formation of neutron-rich heavy nuclei.
Considering the complexity of the reaction mechanism, microscopic approaches could be advantageous in a number of cases. Consequently, several microscopic models have been applied to the $^{238}$U + $^{238}$U collision dynamics, including the Quantum Molecular Dynamics (QMD) model~\cite{TianPRC2008,ZhaoPRC2009,ZhaoPRC2015,ZhaoPRC2013,ZhaoPRC2016}, and the time-dependent Hartree-Fock (TDHF) theory~\cite{GolabekPRL2009,NegeleRMP1982,SimenelEPJA2012,UmarAPPB2018,SekizawaFIP2019,GodbeyFIP2020,SimenelPPNP2018,StevensonPPNP2019}. 
In Ref.~\cite{GolabekPRL2009},  the effect of nuclear quadrupole deformation on the collision time and reaction mechanism has been emphasized. 
One of the most remarkable findings is the occurrence of ternary quasifission in $^{238}$U + $^{238}$U collisions. While fission corresponds to a scission of a compound nucleus with equilibrated intrinsic degrees of freedom, and only depends on its excitation energy and angular momentum, quasifission is a non-equilibrium process in which the composite system of two colliding nuclei splits into fragments, without the formation of a compound system \cite{HindePPNP2021}. The composite system can also split into three fragments, instead of the more commonly observed binary process. 
However, the specific effects of deformation, impact parameter, pairing correlations, and orientation on ternary quasifission are still not well understood \cite{GolabekPRL2009,SimenelPLB2021,HindePRC2022,McGlynnPRC2023,TanakaPRC2023}. We note that the unusual ternary fission mode with a heavy cluster as the third fragment, e.g., Ca or Ni, has also attracted considerable interest \cite{PyatkovEPJA2010,OertzenPLB2014,TashkhodjaevPRC2016}.

In this work, we report a systematic study of ternary quasifission in the collision of $^{238}$U + $^{238}$U, in the microscopic framework of time-dependent covariant density functional theory (TDCDFT). This approach has been successfully developed \cite{RenScpma2019,RenPRC2020,RenPLB2020} and applied to various nuclear phenomena, including alpha-clustering \cite{RenPLB2020}, fission \cite{RenPRL2022,RenPRC2022,LiPRC2023}, and nuclear chirality \cite{RenPRCL2022}.
Static calculations are performed by employing covariant density functional theory (CDFT) in a three-dimensional (3D) lattice space \cite{RenPRC2017,ZhangPRC2022}. To avoid variational collapse, the inverse Hamiltonian method \cite{HaginoPRC2010} is applied, and the Fermion-doubling problem is resolved using the spectral method of Ref.~\cite{RenPRC2017}. Pairing correlations are taken into account dynamically using a monopole interaction in the time-dependent Bardeen-Cooper-Schrieffer (BCS) approximation~\cite{EbataPRC2010,ScampsPRC2013}.

The paper is organized as follows. In Sec.~\ref{sec2}, an outline of the basic formalism of TDCDFT is presented, along with an overview of the time-dependent BCS approximation. The numerical details of the static calculations and time-dependent calculations are included in Sec.~\ref{sec3}. Results are discussed in Sec.~\ref{sec4}, which contains the investigation of effects of octupole deformation, impact parameter, pairing correlations, and orientation. Finally, Sec.~\ref{sec5} concludes the paper with a summary.

\section{Theoretical Framework}\label{sec2}

The time evolution of the single-particle wave function $\psi_k(\bm{r},t)$ is governed by the Dirac equation~\cite{RenPLB2020,EbataPRC2010}
\begin{equation}
	i\hbar\frac{\partial}{\partial t}\psi_k(\bm{r},t)=[\hat{h}(\bm{r},t)-\varepsilon_k(t)]\psi_k(\bm{r},t).
\end{equation}
Here, $\hat{h}(\bm{r},t)$ represents the single-particle Hamiltonian, and $\varepsilon_k(t)=\langle\psi_k(\bm{r},t)|\hat{h}(\bm{r},t)|\psi_k(\bm{r},t)\rangle$ denotes the single-particle energy. 

For the point-coupling relativistic density functional PC-PK1~\cite{ZhaoPRC2010}, the single-particle Hamiltonian $\hat{h}(\bm{r},t)$ can be expressed as:
\begin{equation}
	\hat{h}(\bm{r},t)=\bm{\alpha}\cdot(\hat{\bm{p}}-\bm{V})+V^0+\beta(m+S),
\end{equation}
where
\begin{subequations}
	\begin{eqnarray}
		S(\bm{r},t)&=&\alpha_S\rho_S+\beta_S\rho_S^2+\gamma_S\rho_S^3+\delta_S\Delta \rho_S,\\
		V^{\mu}(\bm{r},t)&=&\alpha_Vj^{\mu}+\gamma_V(j^{\mu}j_{\nu})j^{\mu}+\delta_V\Delta j^{\mu}+\tau_3\alpha_{TV}j_{TV}^{\mu}\nonumber\\
		&+&\tau_3\delta_{TV}\Delta j_{TV}^{\mu}+e\frac{1-\tau_3}{2}A^{\mu},
	\end{eqnarray}
\end{subequations}
$\bm{\alpha}, \beta$ are the Dirac matrices, $m$ is the mass of nucleon, and $\alpha_S, \alpha_V, \alpha_{TV}, \beta_S, \gamma_S, \gamma_V, \delta_S, \delta_V, \delta_{TV}$ are the coupling constants. For further details, we refer the reader to Refs.~\cite{ZhaoPRC2010,RenPRC2020}. The scalar potential $S$ and vector potential $\bm{V}$ are determined by the time-dependent densities and currents as follows:
\begin{subequations}
	\begin{eqnarray}
		\rho_S(\bm{r},t)&=&\sum_k n_k\bar{\psi}_k\psi_k,\\
		j^{\mu}(\bm{r},t)&=&\sum_k n_k\bar{\psi}_k\gamma^{\mu}\psi_k,\\
		j_{TV}^{\mu}(\bm{r},t)&=&\sum_k n_k\bar{\psi}_k\gamma^{\mu}\tau_3\psi_k,
	\end{eqnarray}
\end{subequations}
where $n_k(t)$ represents the occupation probability of the state $k$.

Pairing correlations are considered in the time-dependent BCS approximation, and the wave function of the system can be expressed as:
\begin{equation}
	|\Psi(\bm{r},t)\rangle=\prod_{k>0}[\mu_k(t)+\nu_k(t)c_k^{\dagger}(t)c_{\bar{k}}^{\dagger}(t)]|0\rangle,
\end{equation}
Here, $c_k^{\dagger}(t)$ denotes the creation operator for the single-particle state $\psi_k(t)$, while $c_{\bar{k}}^{\dagger}(t)$ stands for the creation operator of the time-reversed state $\psi_{\bar{k}}(t)$. The parameters $\mu_k(t)$ and $\nu_k(t)$ represent the transformation coefficients between the canonical and quasiparticle states.

The evolution in time of the occupation probability $n_k(t)$, and the pairing tensor $\kappa_k(t)$, are governed by the following equations~\cite{EbataPRC2010,ScampsPRC2013}
\begin{subequations}\label{eq7}
	\begin{eqnarray}
		i\frac{d}{dt}n_k(t)&=&\kappa_k(t)\Delta_k^*(t)-\kappa_k^*(t)\Delta_k(t),\\
		i\frac{d}{dt}\kappa_k(t)&=&[\varepsilon_k(t)+\varepsilon_{\bar{k}}(t)]\kappa_k(t)+\Delta_k(t)[2n_k(t)-1],
	\end{eqnarray}
\end{subequations}
respectively. The gap parameter $\Delta_k(t)$ is determined by the single-particle energies and pairing tensor,
\begin{equation}
	\Delta_k(t)=\left[G\sum_{k'>0}f(\varepsilon_{k'})\kappa_{k'}\right]f(\varepsilon_k),
\end{equation}
where $G$ is the strength parameter of the monopole pairing force, and $f(\varepsilon_k)$ is the cutoff function for the pairing window~\cite{ScampsPRC2013}.

%\subsection{}
%\subsubsection{}

\section{Numerical Details}\label{sec3}
In this work, the density functional PC-PK1~\cite{ZhaoPRC2010} is employed for the particle-hole channel of the effective  interaction. 
The initial states for dynamical calculations are obtained using the self-consistent CDFT method on a 3D-lattice space, with a box size of $L_x\times L_y\times L_z = 24\times 24\times 30~\text{fm}^{3}$. The mesh spacing along each axis is set to $1$ fm. In the dynamical case, the box size is extended to $L_x\times L_y\times L_z = 24\times 24\times 80~\text{fm}^{3}$. To model the time evolution of single-particle wave functions, a predictor-corrector method is utilized, with a fourth-order Taylor expansion of the time-evolution operator. The time step is chosen $6.67\times 10^{-25}$ s. The pairing window cutoff function used in this work follows the formula included in Ref.~\cite{RenPRC2022}. The pairing strength parameters are determined by the empirical gaps using the three-point odd-even mass formula. In the case of $^{238}$U, a pairing strength of $G_n = -0.135$ MeV is assigned to neutrons, while a pairing strength of $G_p = -0.230$ MeV is utilized for protons. For $^{226}$Ra, the pairing strengths $G_n = -0.155$ MeV and $G_p = -0.250$ MeV are used for neutrons and protons, respectively. Although the time-dependent BCS approximation violates the one-body continuity equation~\cite{ScampsPRC2012}, its effect is found to be moderate in this work. The particle number in each fragment changes only slightly after scission.

\section{Results and discussion}\label{sec4}
\subsection{Octupole deformation effects}

\begin{figure}[!htbp]
	\centering
	\includegraphics[width=0.9\textwidth]{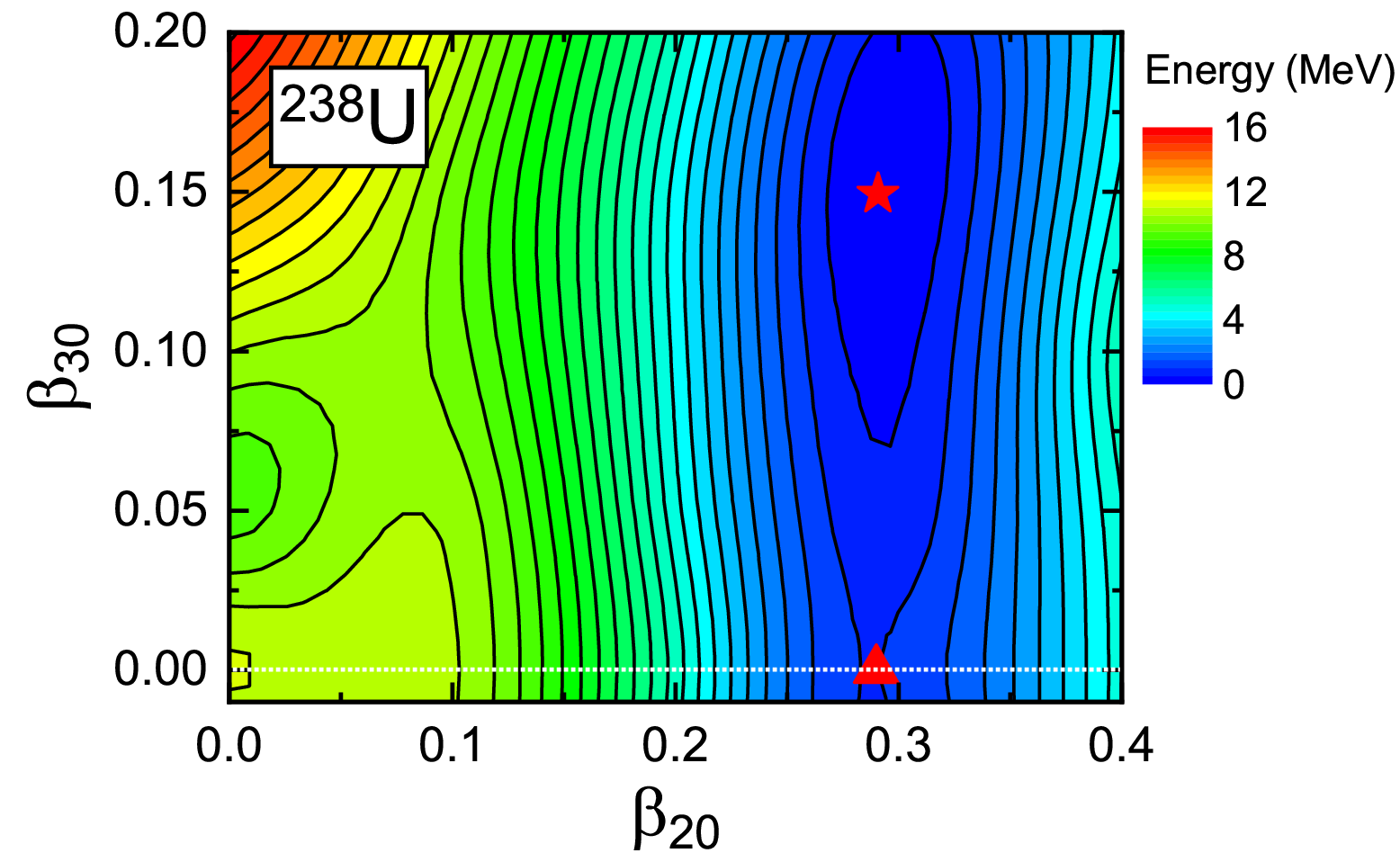}
	\caption{(Color online) The deformation energy surface of $^{238}$U in the $(\beta_{20},\beta_{30})$ plane, calculated with the CDFT on a 3D-lattice space. Here pairing correlations are not included. Neighbouring contours on the surface differ in energy by $0.5$ MeV. The equilibrium minimum is denoted by the star symbol, while the energy minimum along the $\beta_{30} = 0$ direction is indicated by the triangle.}
	\label{fig1}
\end{figure}

Figure~\ref{fig1} displays the deformation energy surface of $^{238}$U in the $(\beta_{20},\beta_{30})$ plane, obtained using the 3D-lattice CDFT with the PC-PK1 interaction but, for the moment, without the inclusion of pairing correlations. The equilibrium minimum is denoted by the star symbol, while the energy minimum along the $\beta_{30} = 0$ direction is indicated by a triangle symbol. The energy surface exhibits pronounced softness along the $\beta_{30}$ direction near the global minimum. The deformation parameters of the equilibrium minimum are: $\beta_{20} = 0.29$ and $\beta_{30} = 0.15$. When reflection symmetry is imposed, the energy minimum still occurs at $\beta_{20} = 0.29$, but at $0.99$ MeV higher energy.

Typically, the ground state of $^{238}$U is chosen as the initial state for time-dependent calculations. In order to investigate the influence of octupole deformation on ternary quasifission dynamics, we consider two initial states: the equilibrium minimum (the star in Fig.~\ref{fig1}), and the quadrupole prolate state (the triangle in Fig.~\ref{fig1}).

\begin{figure*}[!htbp]
	\centering
	\includegraphics[width=0.8\textwidth]{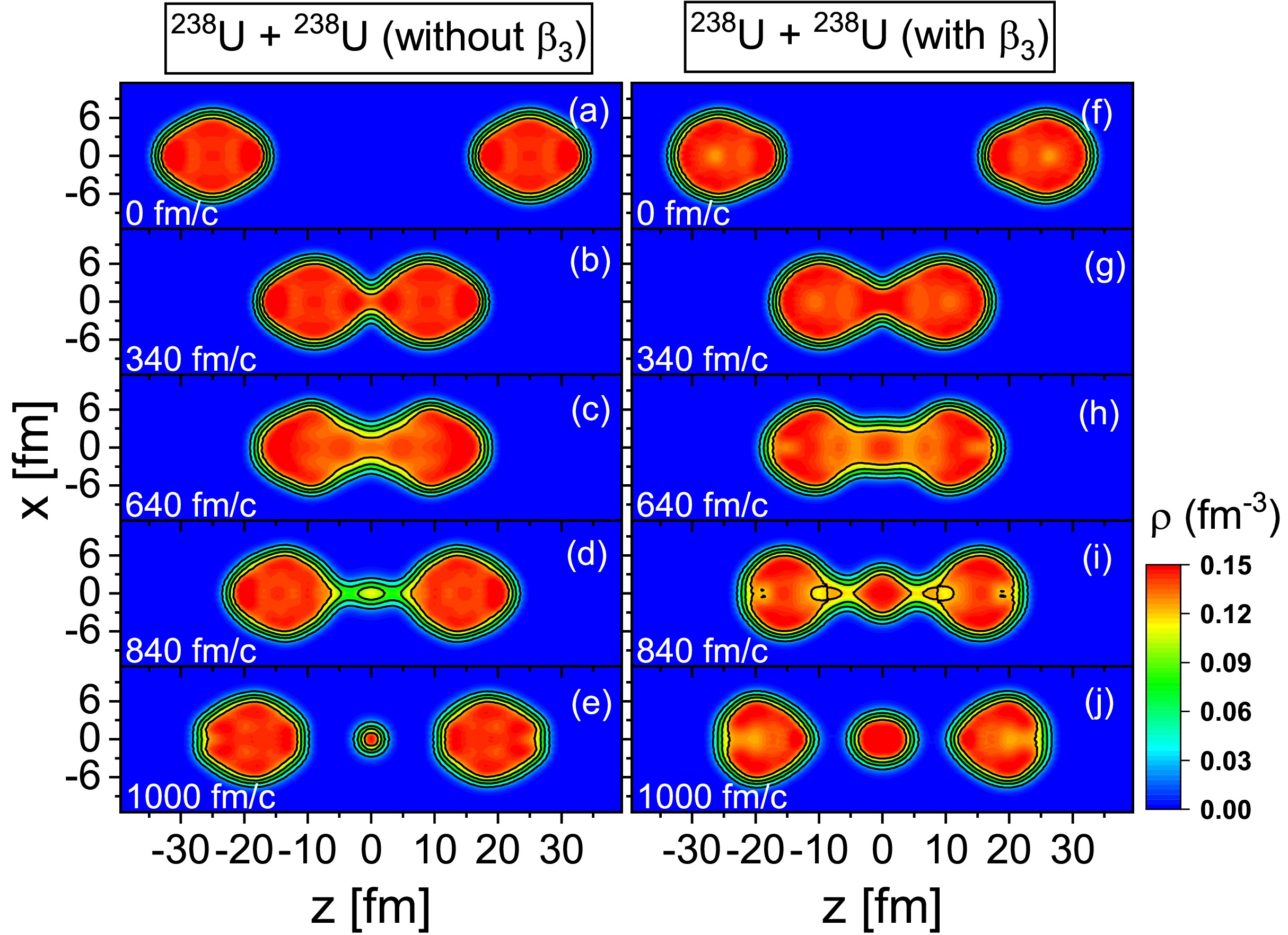}
	\caption{(Color online) The total density in the $x$-$z$ plane at the times $t = 0,~340,~640,~840,\text{and}~1000$ fm/c for the central collision of $^{238}$U + $^{238}$U at a center-of-mass energy $E_{\text{c.m.}}=900$ MeV, calculated using  the TDCDFT. In the left column, the initial states of $^{238}$U are quadrupole prolate deformed local minima with no octupole degree of freedom. In the right column, the initial states correspond to the equilibrium minimum of $^{238}$U with an octupole deformation $\beta_{30} = 0.15$.}
	\label{fig2}
\end{figure*}

Figure~\ref{fig2} illustrates the time evolution of the total density in the $x$-$z$ plane at specific times: $t = 0,~340,~640,~840,\text{and}~1000$ fm/c for the central head-to-head collision of $^{238}$U + $^{238}$U, at a center-of-mass energy of $E_{\text{c.m.}}=900$ MeV. The left column of the figure corresponds to a scenario with quadrupole prolate initial states and no octupole degree of freedom, whereas in the right column snapshots of the total density are plotted that display the time evolution for initial states with the octupole moment of the equilibrium minimum.   

In the collision of $^{238}$U + $^{238}$U, the initial orientations are not unique due to the presence of a deformed initial state, as discussed in Ref.~\cite{GolabekPRL2009}. In Fig.~\ref{fig2} we consider the head-to-head orientation. For the specific choice of collision energy and initial distance, the two $^{238}$U nuclei come into contact at approximately $t = 340$ fm/c (Fig.~\ref{fig2}(b) and (g)). A composite system is formed that does not equlibrate, rather it dissociates into fragments after a certain period of time. As we follow the time evolution to $t = 640$ fm/c, a neck has already formed in both cases, as shown in Fig.~\ref{fig2}(c) and (h). Notably, the neck is significantly thicker and longer when the initial state exhibits octupole deformation. This already indicates the influence of octupole deformation on the dynamics of collision.

At $t = 840$ fm/c, the system approaches the breaking point, with a middle fragment forming in the neck. A significant difference between the densities of the middle fragment is observed in Figs.~\ref{fig2}(d) and (i). The enhanced density in Fig.~\ref{fig2}(i) contributes to the formation of a larger middle fragment, as shown in Figs.~\ref{fig2}(e) and (j). When only quadrupole deformation is considered, the middle fragment exhibits the average number of protons and neutrons of $^{13}$B. On the other hand, when also the octupole degree of freedom is taken into account, the middle fragment corresponds to $^{55}$Ca. This result indicates that the presence of octupole deformation will generally have a pronounced effect on the number of protons and neutrons of the middle fragment that is formed in ternary quasifission of actinide nuclei. 

\begin{figure}[!htbp]
	\centering
	\includegraphics[width=0.9\linewidth]{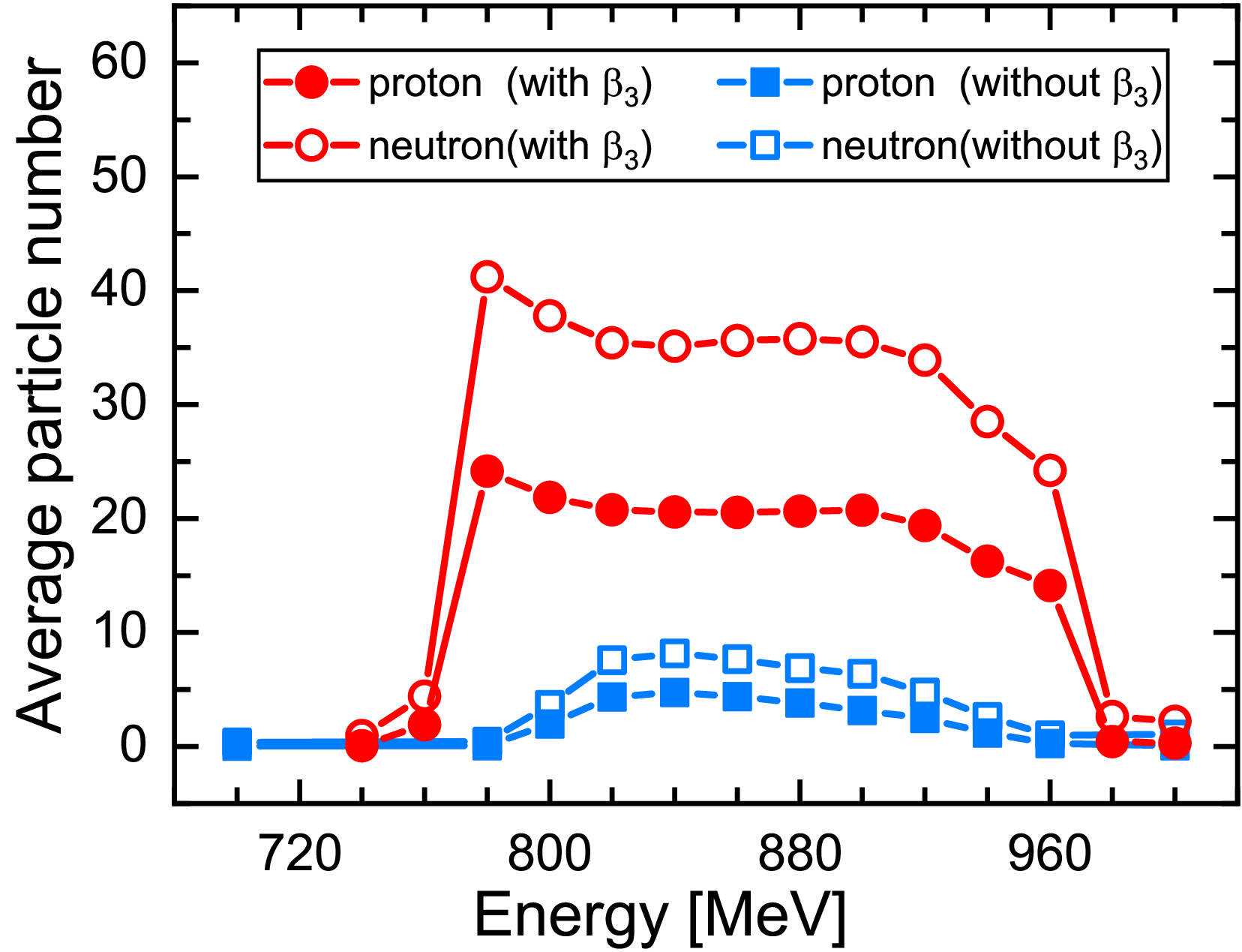}
	\caption{(Color online) The average number of protons and neutrons in the middle fragment for the $^{238}$U + $^{238}$U collision, as a function of the center-of-mass energy. The filled (empty) squares represent the protons (neutrons). The colour red (blue) corresponds to the collision with (without) an octupole-deformed initial state.}
	\label{fig3}
\end{figure}

Figure \ref{fig3} displays the average number of protons and neutrons in the middle fragment as a function of the center-of-mass energy. The filled (empty) squares represent the protons (neutrons), while the colour red (blue) corresponds to a collision with (without) an octupole-deformed initial state. 

When the center-of-mass energy is lower than 780 MeV, ternary quasifission does not take place. As the energy increases, ternary fragments begin to emerge. The occurrence of ternary quasifission is limited to a rather narrow range of energies. In the absence of octupole deformation, this energy window spans the interval from 800 MeV to 920 MeV. When the initial state has octupole deformation, the energy window for ternary quasifission in head-to-head collisions extends from 780 MeV to 960 MeV. Along the whole interval in which ternary quasifission occurs, the average number of protons and neutrons in the middle fragment is significantly larger when the octupole degree of freedom is taken into account. 

\begin{figure}[!htbp]
	\centering
	\includegraphics[width=\linewidth]{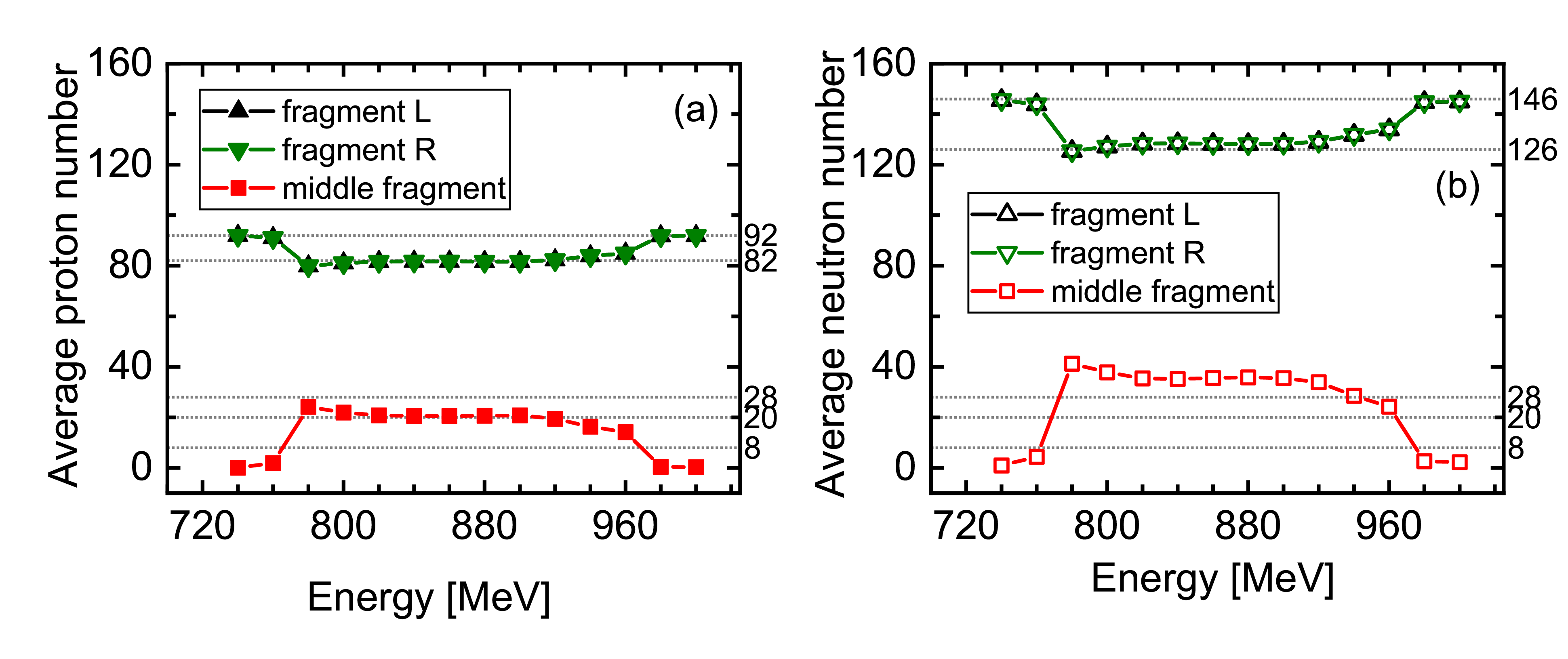}
	\caption{(Color online) The average number of protons (left panel) and neutrons (right panel) in the three fragments for an $^{238}$U + $^{238}$U collision, with the octupole-deformed equilibrium initial state, as functions of the center-of-mass energy. The nucleus evolving in the $z < 0 $ region is labelled as fragment L, and the one in the $z > 0 $ plane is fragment R.}
	\label{fig4}
\end{figure}

One also notices that, with the octupole-deformed equilibrium as initial state, the average particle numbers of the middle fragment change rather abruptly at both ends of the energy interval in which ternary quasifission takes place. In Fig.~\ref{fig4}, we plot the average number of protons and neutrons in the three reaction products. The nucleus located in the $z < 0$ region is denoted as fragment L, and the nucleus in the $z > 0$ region is fragment R. In Fig.~\ref{fig4} (a) and (b), the average particle numbers of fragment L and fragment R are identical, and the entire system exhibits reflection symmetry. At energy $E_{\text{c.m.}} = 780$ MeV, at which the ternary quasifission appears, the average proton number suddenly changes from $Z = 92$ to $Z = 80$, while the number of neutron decreases from $N = 146$ to $N = 126$. At the same time, the remaining nucleons form the middle fragment, characterized by the average proton and neutron number  $Z = 24$ and $N = 40$, respectively. Within the energy interval from $E = 780$ MeV to $E = 960$ MeV, the average proton number in fragments L and R ranges from $Z = 80$ to $Z = 85$, while the average neutron number varies from $N = 126$ to $N = 134$. We note that the proton numbers in the heavy fragments align closely with the line $Z = 82$, while the number of neutrons is close to $N = 126$. This suggests a scenario in which the two $^{238}$U nuclei retain the core of $^{208}$Pb, while the remaining constituents form a neutron-rich middle fragment. As the energy increases beyond $E = 960$ MeV, ternary quasifission does not take place any more, and the average particle numbers of fragment L and fragment R are again $Z = 92$ and $N = 146$, respectively. Therefore, it appears that the shell closures at $Z = 82$ and $N = 126$ exerts a significant influence on the transfer of nucleons in ternary quasifission, as it has previously been suggested in studies based on TDHF \cite{WakhlePRL2014,SimenelPLB2021,HindePRC2022,McGlynnPRC2023}.

\subsection{Effect of the impact parameter}

\begin{figure}[h]
	\centering
	\includegraphics[width=\linewidth]{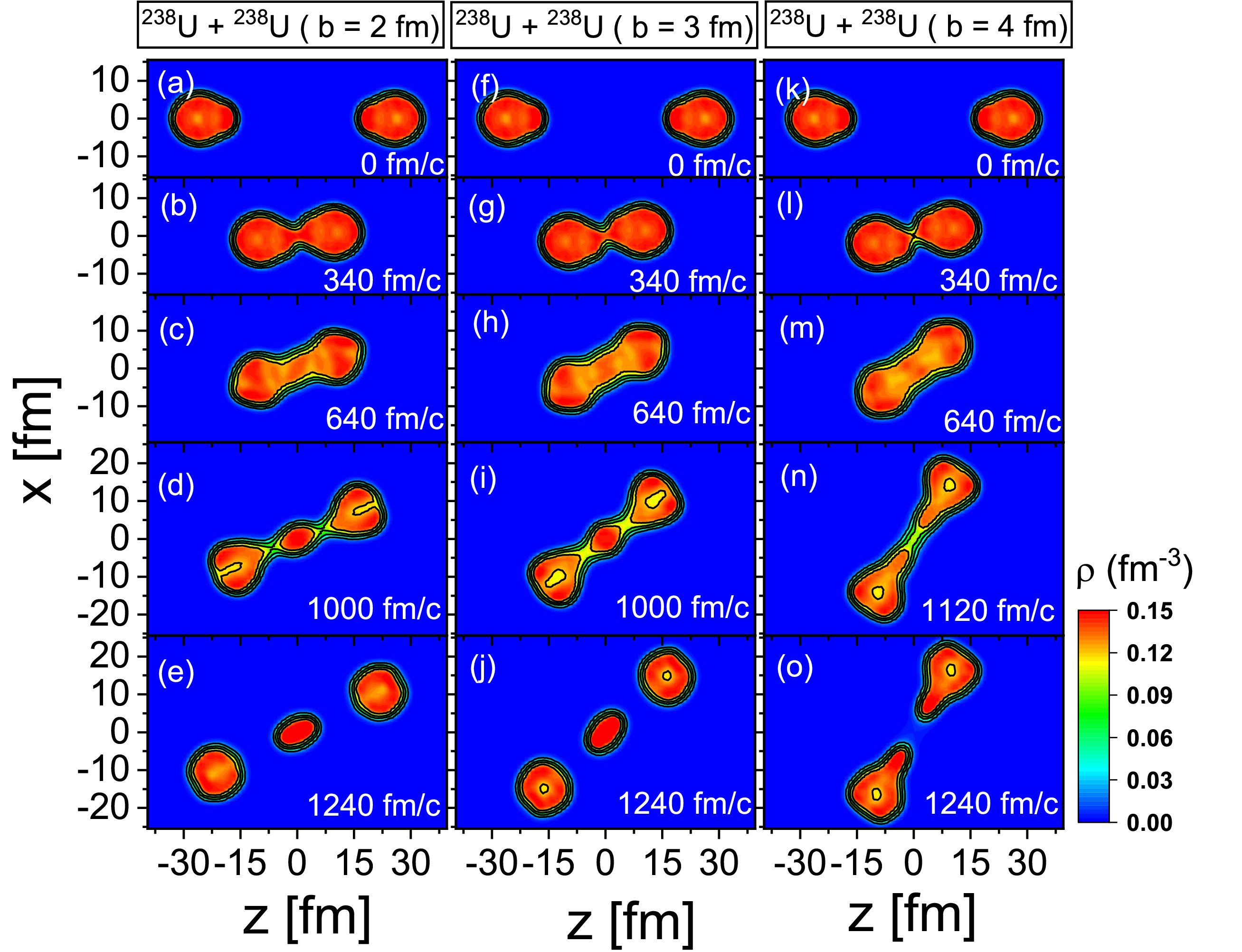}
	\caption{(Color online) Total densities in the $x$-$z$ plane for the $^{238}$U + $^{238}$U collision at the center-of-mass energy $E_{\text{c.m.}}=900$ MeV, with an initial head-to-head orientation. The columns from left to right correspond to impact parameters of $b = 2$ fm, $b = 3$ fm, and $b = 4$ fm, respectively. Density snapshots are shown at times $t = 0,~340,~640,~1000,\text{ and }1240$ fm/c for $b = 2$ fm and $b = 3$ fm, while at times $t = 0,~340,~640,~1120,\text{ and }1240$ fm/c for $b = 4$ fm.}
	\label{fig5}
\end{figure}

In the preceding section, we have considered only the case of central collisions, with the impact parameter set to zero. To calculate cross sections for ternary quasifission, one must also take into account collisions with varying values of the impact parameter. As an illustration, Figure~\ref{fig5} displays the distribution of the total density in the $x$-$z$ plane for the head-to-head $^{238}$U + $^{238}$U collision at the center-of-mass energy $E_{\text{c.m.}}=900$ MeV. Three values of the impact parameter are considered, namely $b = 2$ fm (left column), $b = 3$ fm (middle column), and $b = 4$ fm (right column). In all three cases, the contact between the two $^{238}$U nuclei occurs approximately at $t = 340$ fm/c, as shown in Figs.~\ref{fig5}(b), (g), and (l). The neck of the fissioning system is formed at $t \approx 640$ fm/c, with the thickness of the neck increasing and the length decreasing as the impact parameter becomes larger (Figs.~\ref{fig5}(c),(h), and (m)). As the system approaches scission, neck density variations are observed in the central region, as seen in Figs. \ref{fig5}(d), (i), and (n). Notably, when the impact parameter reaches $b = 4$ fm and beyond, the density in the central region decreases significantly, and the evolution time of the neck appears longer. Consequently, when the system is at scission, the particles in the neck are absorbed by the fragments on both sides, and the density in the central region cannot support the formation of a middle fragment (Fig.~\ref{fig5}(o)). 

\begin{figure}[h]
	\centering
	\includegraphics[width=0.8\linewidth]{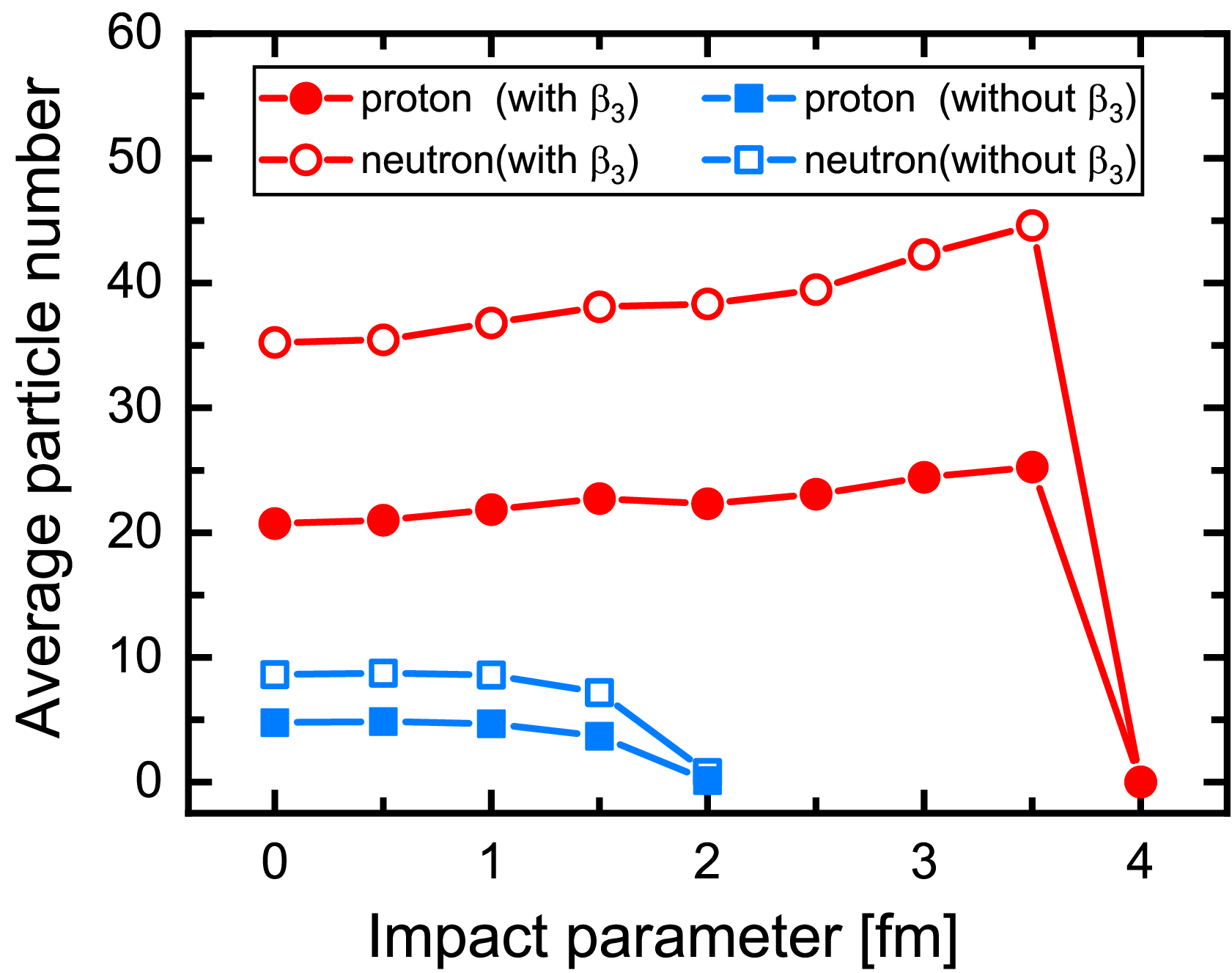}
	\caption{(Color online) The average number of protons and neutrons in the middle fragment for the $^{238}$U + $^{238}$U collision at the center-of-mass energy $E_{\text{c.m.}}=900$ MeV, as functions of the impact parameter. The filled (empty) squares denote protons (neutrons). The colour red (blue) corresponds to the collision with (without) an octupole-deformed initial state.}
	\label{fig6}
\end{figure}

The average number of particles in the middle fragment is also shown in Fig.~\ref{fig6}, including the case of a quadrupole prolate initial state with no octupole degree of freedom. In the case of a quadrupole initial state, the particle numbers of the middle fragment exhibit small variations when the impact parameter is smaller than $1.5$ fm, corresponding to a $^{13}$B-like fragment. As the impact parameter increases to $1.5$ fm, the calculated middle fragment is more like  $^{11}$Be, and finally ternary quasifission does not occur any longer when the impact parameter is equal or greater than $2$ fm. On the other hand, for the initial state with octupole deformation, the average particle numbers gradually increase, ranging from $21 \leq Z \leq 25$ and $35 \leq N \leq 45$, as the impact parameter increases from $b = 0$ fm to $3.5$ fm. Once the impact parameter reaches $b = 4$ fm, ternary quasifission no longer takes place. The inclusion of octupole deformation allows for a wider interval of impact parameters for which ternary quasifission is predicted. Consequently, the cross section is expected to increase when octupole deformation is taken into account. 

\subsection{Pairing effects}

\begin{figure}[h]
	\centering
	\includegraphics[width=\linewidth]{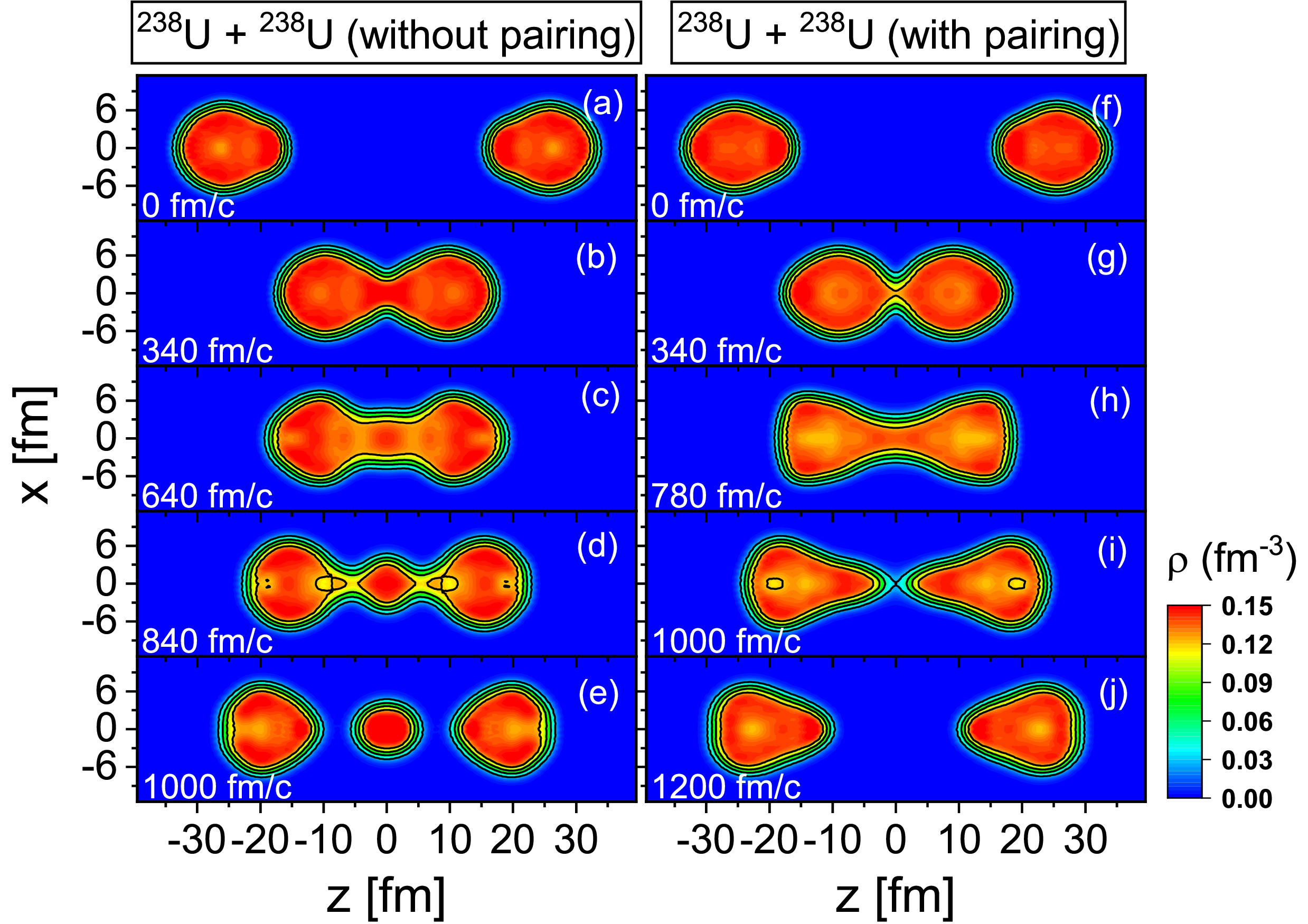}
	\caption{(Color online) Total densities in the $x$-$z$ plane are for the central collision of $^{238}$U + $^{238}$U at the center-of-mass energy $E_{\text{c.m.}}=900$ MeV, with an initial head-to-head orientation. The left column shows the densities at times $t = 0,~340,~640,~840,\text{ and }1000$ fm/c, calculated without inclusion of pairing correlations. The right column displays the densities at times $t = 0,~340,~780,~1000,\text{ and }1200$ fm/c, with pairing treated dynamically in the time-dependent BCS approximation.}
	\label{fig7}
\end{figure}

In the next step, we include pairing correlations in the time-dependent BCS approximation, and analyze their effect on ternary quasifission dynamics. In the calculation of the deformation energy surface of $^{238}$U, the inclusion of pairing correlations shifts the equilibrium minimum to a smaller octupole deformation $\beta_{30} = 0.08$, compared to the result obtained without pairing 
($\beta_{30} = 0.15$, cf. Fig.~\ref{fig1}). In Fig.~\ref{fig7}, we compare snapshots of total densities  in the $x$-$z$ plane for the central collision of $^{238}$U + $^{238}$U at the center-of-mass energy $E_{\text{c.m.}}=900$ MeV, with an initial head-to-head orientation, calculated without (left) and with (right) pairing correlations. The effect of pairing is, obviously, very pronounced. While without pairing one finds a third fragment forming in the neck of the fissioning system, it appears that the inclusion of pairing prevents the occurrence of ternary quasifission. We have verified that, with pairing included, ternary quasifission does not take place for energies in the interval between 780 MeV and 960 MeV, which corresponds to the energy window for ternary quasifission in the $^{238}$U + $^{238}$U collision with an octupole-deformed initial state. 

Given the importance of the octupole degree of freedom, and the shell closures at $Z = 82$ and $N = 126$, for the ternary quasifission mechanism, we have also considered the collision of two $^{226}$Ra nuclei. The choice of $^{226}$Ra is motivated by its distinct characteristics, including a deformation energy surface that exhibits a more pronounced octupole deformation  compared to $^{238}$U. Without pairing, the equilibrium minimum for $^{226}$Ra is calculated at $\beta_{20} = 0.21$ and $\beta_{30} = 0.16$. When the monopole pairing interaction is included with the strength parameters determined by the empirical pairing gaps, the deformation parameters of the equilibrium minimum are: $\beta_{20} = 0.20$ and $\beta_{30} = 0.13$. 

\begin{figure}[h]
	\centering
	\includegraphics[width=\linewidth]{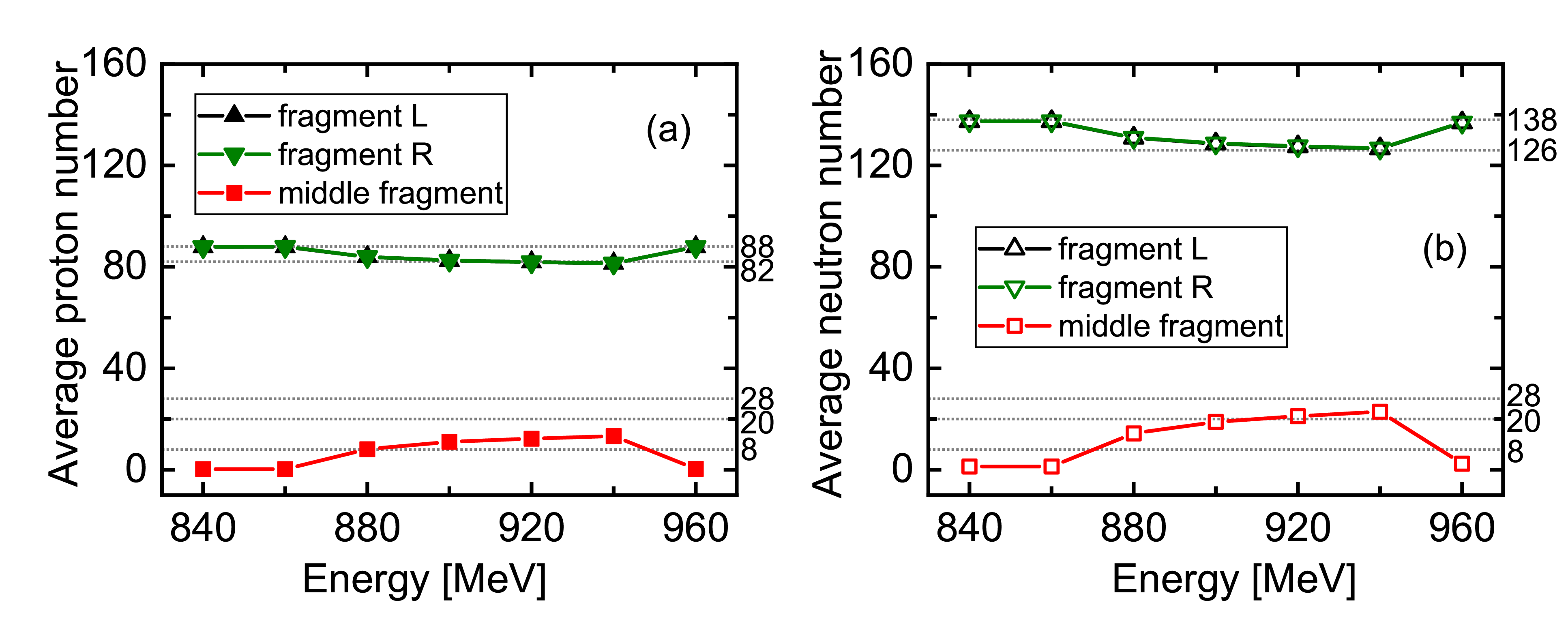}
	\caption{(Color online) Same as Fig. \ref{fig4}, but for the $^{226}$Ra + $^{226}$Ra collision.}
	\label{fig8}
\end{figure}

Figure~\ref{fig8} displays the average number of protons (left panel) and neutrons (right panel) in the fragments resulting from $^{226}$Ra + $^{226}$Ra head-to-head collisions, calculated without pairing, as a function of center-of-mass energy. The particle numbers in the middle fragment show similarities to those calculated in the case of $^{238}$U + $^{238}$U collisions, though in a narrower energy interval between $880$ MeV and $940$ MeV. Within this energy window, the average number of neutrons in fragments L or R varies between $N = 126$ and $N = 131$, while the average number of protons ranges from $Z = 81$ to $Z = 84$. The middle fragment, characterized by $8 \leq Z \leq 13$ and $14 \leq N \leq 23$, exhibits a higher neutron-to-proton ratio compared to the $^{238}$U + $^{238}$U case. One notices that the average number of protons and neutrons in fragments L and R are concentrated along the lines $Z = 82$ and $N = 126$, respectively, just as in  the case of $^{238}$U + $^{238}$U collisions, emphasizing the role of shell effects in ternary quasifission. 
\begin{figure}[h]
	\centering
	\includegraphics[width=\linewidth]{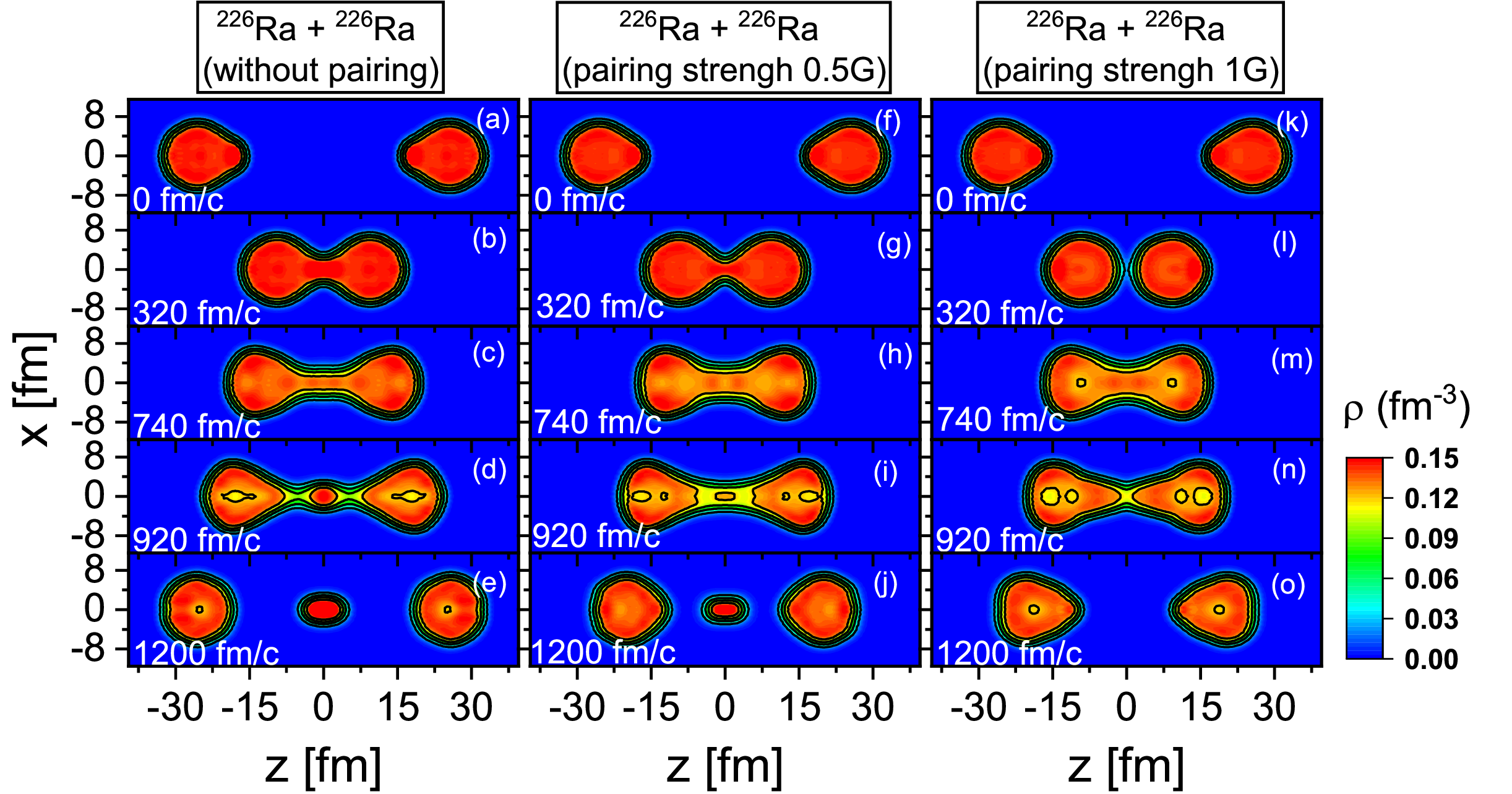}
	\caption{(Color online) Total densities in the $x$-$z$ plane, at specific instants $t = 0,~320,~740,~920,~\text{and}~1200$ fm/c, for the central, head-to-head collision $^{226}$Ra + $^{226}$Ra at the center-of-mass energy $900$ MeV. In the left column, the pairing interaction is not included. In the right column pairing is taken into account using the time-dependent BCS approximation with the strength parameters $(G_n, G_p) = (-0.155, -0.250)$ MeV determined from empirical pairing gaps, while in the middle column the pairing strengths are reduced by 50\%: $(0.5G_n, 0.5G_p)$.}
	\label{fig9}
\end{figure}

To illustrate the impact of pairing correlations on ternary quasifission also in the case of $^{226}$Ra nuclei, in Fig.~\ref{fig9} we plot the total densities in the $x$-$z$ plane at times $t = 0,~320,~740,~920, \text{and}~1200$ fm/c for the central, head-to-head collision of $^{226}$Ra + $^{226}$Ra, at the center-of-mass energy $900$ MeV. For comparison, the left column displays the results obtained when the pairing interaction is not included, in the right column the pairing strengths $(G_n, G_p) = (-0.155, -0.250)$ MeV are determined from empirical pairing gaps, while in the middle column the pairing strengths are reduced by a factor 2: $(0.5G_n, 0.5G_p)$. We note that, in the latter case, the pairing energy is nearly zero.

The initial orientation of the two $^{226}$Ra nuclei is shown in Figs.~\ref{fig9}(a), (f), and (k). They come into contact at approximately $t = 320$ fm/c. Subsequent to the formation of the composite system, a neck of the fissioning nucleus can be observed at $t \approx 740$ fm/c (Figs.~\ref{fig9}(c), (h), and (m)). One notices that the neck appears thicker and shorter when pairing correlations are included, and the density in the middle decreases as the pairing strength increases. As shown  in Fig.~\ref{fig9}(n), the nucleon density in the central region is reduced as the neck elongates, and the particles in the neck are eventually absorbed by the two fragments. Ternary quasifission does not occur in the case of full pairing strength (Fig.~\ref{fig9}(o)), but in the intermediate case, when pairing is reduced, a relatively small middle fragment appears at scission (Fig.~\ref{fig9}(j)). 

\subsection{Orientation effects}

\begin{figure}[h]
	\centering
	\includegraphics[width=\linewidth]{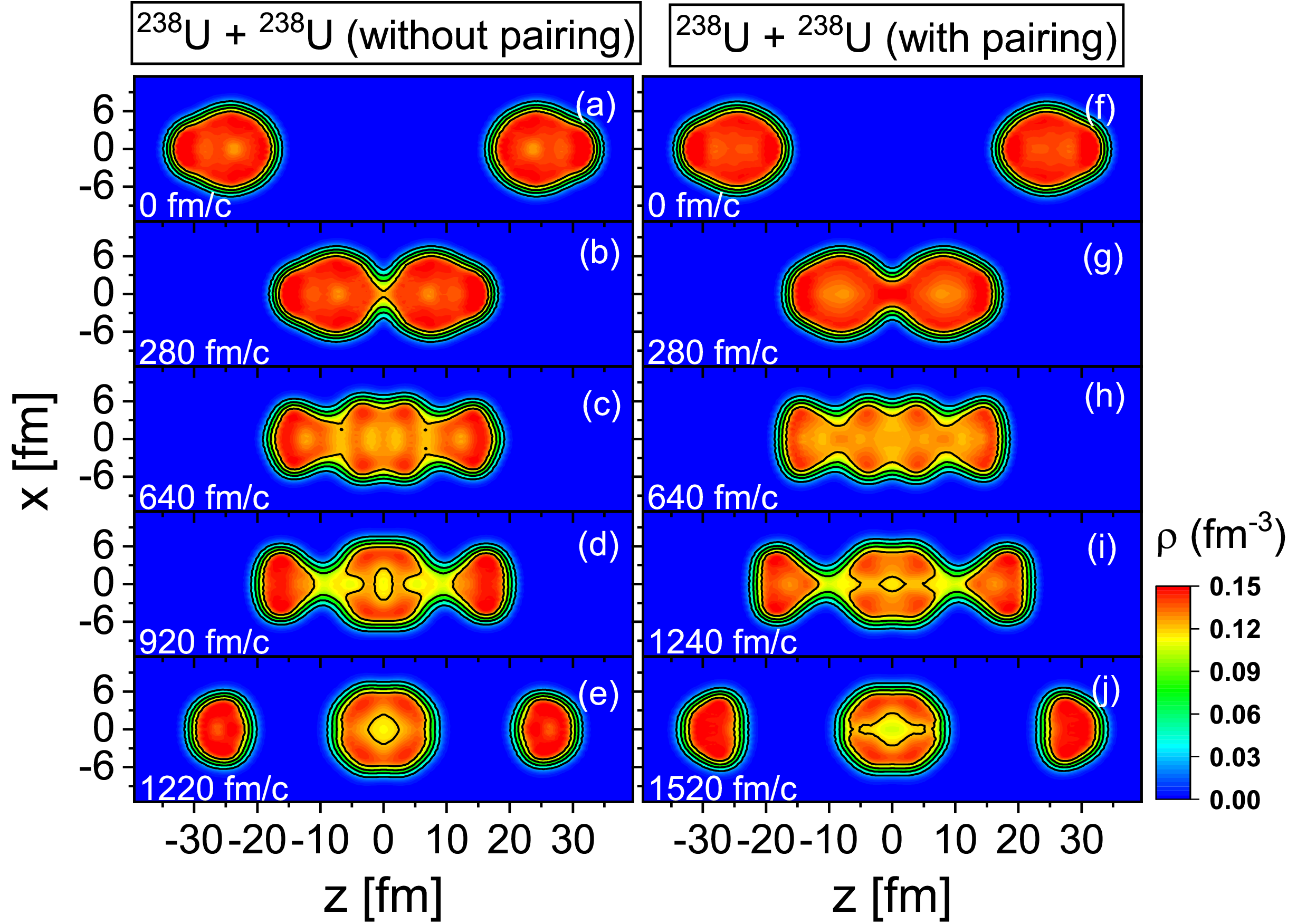}
	\caption{(Color online) Total densities in the $x$-$z$ plane for the central collision of $^{238}$U + $^{238}$U at the center-of-mass energy $E_{\text{c.m.}}=1200$ MeV, with an initial tail-to-tail orientation. The left column shows the densities at times $t = 0,~280,~640,~920,\text{ and }1220$ fm/c, without the inclusion of pairing correlations. The right column displays the densities at times $t = 0,~280,~640,~1240,\text{ and }1520$ fm/c, taking into account the pairing monopole interaction with empirical strength parameters.}
	\label{fig10}
\end{figure}

The inclusion of pairing correlations appears to prevent the formation of a middle fragment in head-to-head collisions, both in the cases of $^{238}$U and $^{226}$Ra nuclei. It is, therefore, of interest to consider different orientations. 
Figure~\ref{fig10} displays the time evolution of the total nucleon density distribution in the $^{238}$U + $^{238}$U collision at the center-of-mass energy $E_{\text{c.m.}}=1200$ MeV, with an initial tail-to-tail orientation. And even though there are noticeable differences in the evolution of densities calculated without and with pairing, in both cases a heavy middle fragment is formed at scission.

\begin{figure}[h]
	\centering
	\includegraphics[width=0.8\linewidth]{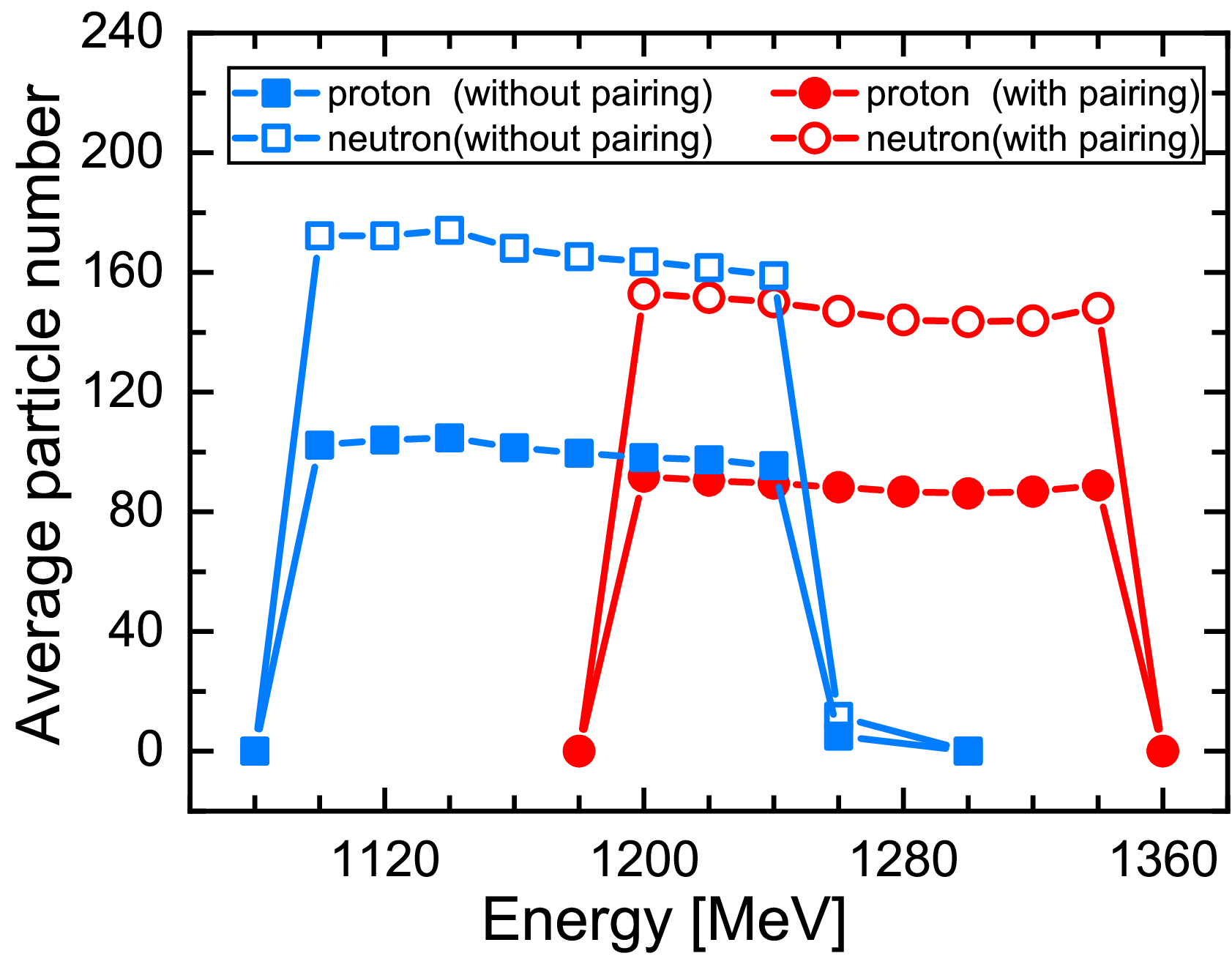}
	\caption{(Color online) The average number of protons and neutrons in the middle fragment for central $^{238}$U + $^{238}$U collisions with an initial tail-to-tail orientation, are shown as functions of the center-of-mass energy. The filled (empty) squares denote protons (neutrons). The colour red (blue) corresponds to the calculation with (without) pairing correlations.}
	\label{fig11}
\end{figure}

The average proton and neutron numbers of the middle fragment are plotted in Fig.~\ref{fig11}, as functions of the center-of-mass energy. While for this orientation, there is not so much difference in the number of protons and neutrons in the middle fragment when calculated without and with pairing, an interesting effect is that the inclusion of pairing shifts the interval in which ternary quasifission occurs to higher energies by more than 100 MeV. In the absence of pairing, the average nucleon numbers of the third fragment are calculated in the interval $95\leq Z\leq 105$ and $159\leq N\leq174$. When pairing correlations are taken into account, the average particle numbers change to $86\leq Z\leq 92$ and $144\leq N\leq153$. This means that neutron-rich nuclei in the uranium region could be produced as the middle fragment in the tail-to-tail orientation for central collisions.

\begin{figure}[h]
	\centering
	\includegraphics[width=\linewidth]{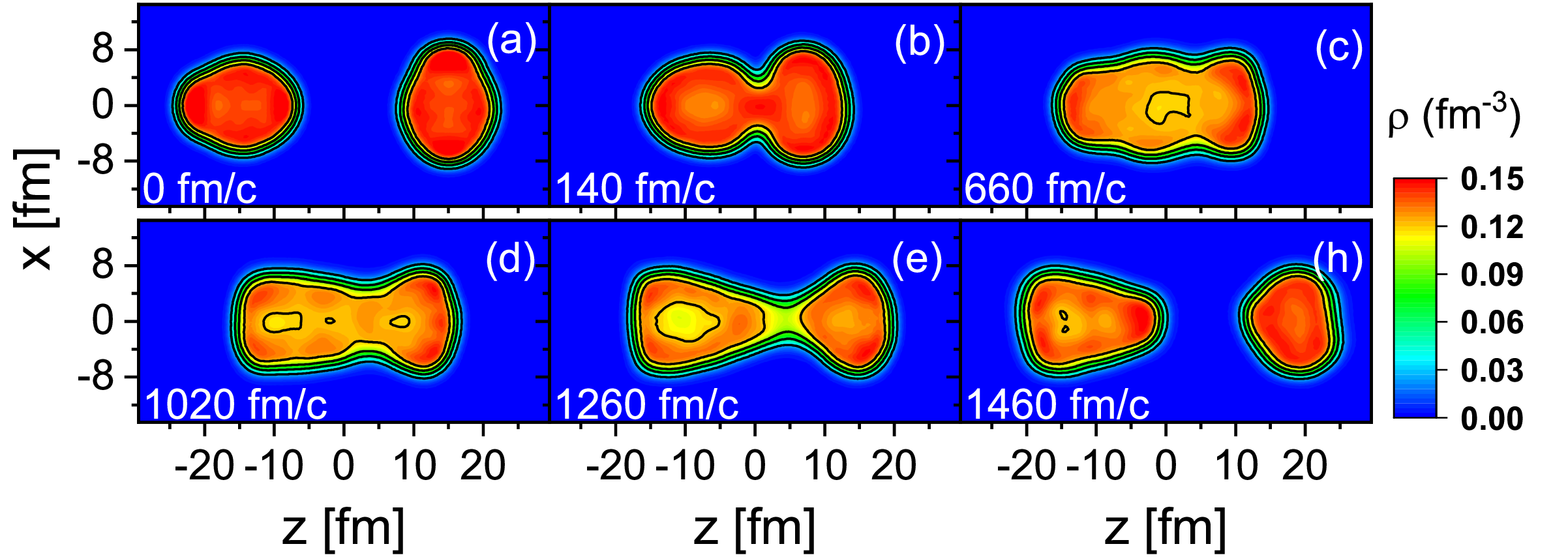}
	\caption{(Color online) The total nucleon density in the $x$-$z$ plane at times ($t = 0,~140,~660,~1020,~1260,~\text{and}~1460$ fm/c ) for the central collision of $^{238}$U + $^{238}$U at the center-of-mass energy of $1300$ MeV, and the tail-to-side initial orientation. The pairing correlations are taken into account in the calculations.}
	\label{fig12}
\end{figure}

Finally, in Fig.~\ref{fig12} we show the total nucleon density evolution in the $x$-$z$ plane for the central $^{238}$U + $^{238}$U collision at the center-of-mass energy of $1300$ MeV, and the tail-to-side initial orientation. The pairing interaction is included in the calculation. In panel (a) of Fig.~\ref{fig12}, the two $^{238}$U nuclei are placed at an initial separation distance of $30$ fm. This is a somewhat smaller distance than in the head-to-head and tail-to-tail cases, and is due to the fact that the lattice had to be extended in the perpendicular $x$-direction. At $t = 140$ fm/c, shown in panel (b), the nuclei come into contact. Subsequently, the density of the composite system exhibits oscillations, as observed in panels (c) and (d). As time evolves to $1260$ fm/c, the entire system reaches a critical point when it splits into two fragments, as shown in panel (e). During the scission process, the majority of nucleons are transferred to the heavier fragment, resulting in an average particle number of $Z = 113$ and $N = 182$ for the heavy fragment. No ternary quasifission occurs for the tail-to-side initial orientation.

\begin{figure}[h]
	\centering
	\includegraphics[width=\linewidth]{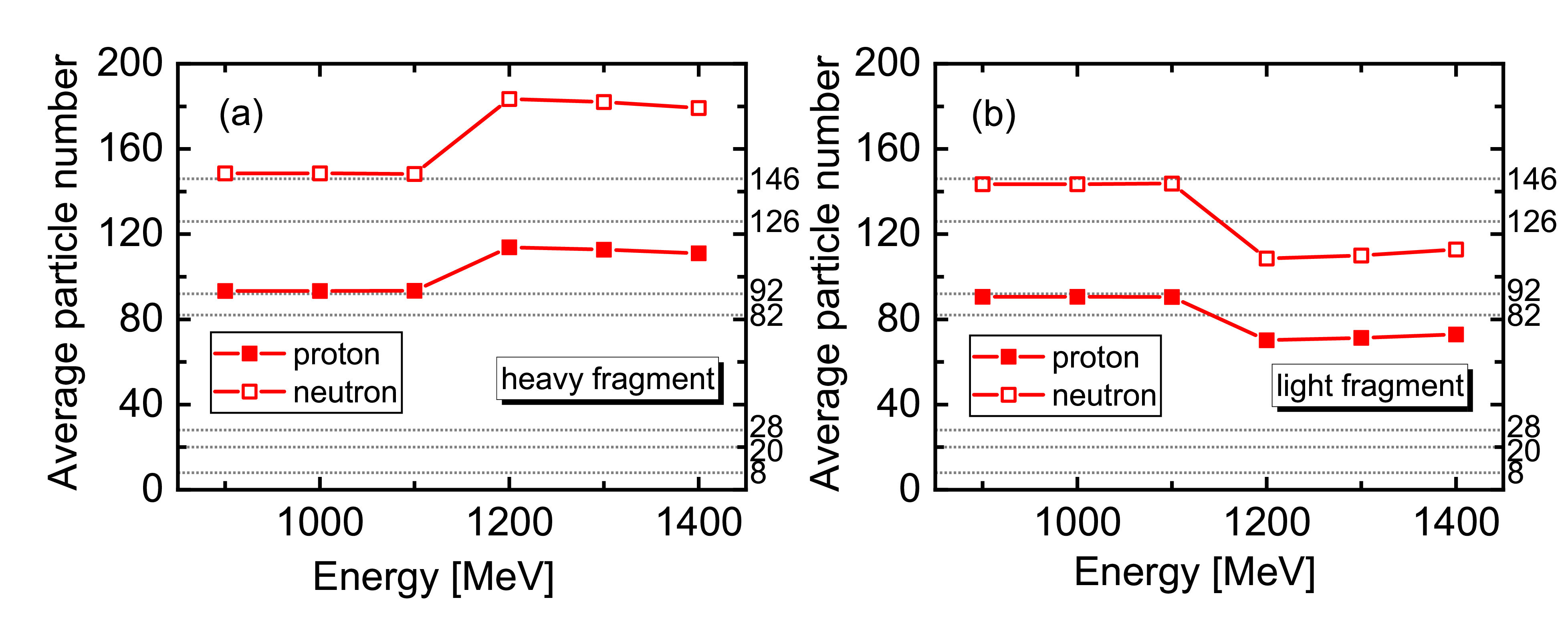}
	\caption{(Color online) The average number of protons and neutrons in the heavy (left), and light (right) fragments as functions of the center-of-mass energy for the central $^{238}$U + $^{238}$U collision, with tail-to-side initial orientation.}
	\label{fig13}
\end{figure}

Because of this very interesting result, we have analyzed the average contents of the heavy and light fragments for the tail-to-side central $^{238}$U + $^{238}$U collision in the energy interval from $900$ MeV to $1400$ MeV. The average number of protons and neutrons are plotted in Fig.~\ref{fig13}. At lower energies, ranging from $E = 900$ to $1100$ MeV, the transfer of nucleons is very limited, and the heavy fragment contains on average $Z = 93$ protons and $N = 149$ neutrons, while the light fragment consists of $Z = 91$ and $N = 143$ nucleons. As the energy increases, a larger number of nucleons is transferred, resulting in the formation of neutron-rich heavy nuclei. For energies above $1100$ MeV, the proton number of the heavy fragment varies from $111 \leq Z \leq 114$, and the corresponding neutron number is in the range $180 \leq N \leq 183$. It appears that multi-nucleon transfer predicted in the tail-to-tail and tail-to-side orientations, presents a feasible mechanism for the production of the heaviest neutron-rich nuclei.

\section{Summary}\label{sec5}

Collisions of pairs of $^{238}$U nuclei have systematically been analyzed using the microscopic framework of TDCDFT, that dynamically includes pairing correlations. In particular, by considering $^{238}$U + $^{238}$U collisions at various energies and impact parameters, we have investigated the process of ternary quasifission of the composite system. Considering also the case of two $^{226}$Ra nuclei it has been shown that, in addition to quadrupole deformation, the inclusion of octupole degrees of freedom has a pronounced effect on the formation of a middle fragment at scission. The formation of a much larger fragment, and in a broader interval of center-of-mass energies is predicted when an octupole deformation of the initial equilibrium state of the colliding nuclei is taken into account. Non-central collisions with different impact parameters have also been considered and, as one would expect, the occurrence of ternary quasifission is inhibited as the impact parameter increases beyond a critical value. Notably, the inclusion of octupole deformation plays a significant role in extending the range of impact parameters in which ternary quasifission can take place, and this will result in a significant contribution to the cross-section for ternary quasifission. 

The effect of pairing correlations has been investigated by employing a monopole pairing force in the time-dependent BCS approximation. In the case of head-to-head collisions, both for the $^{238}$U and $^{226}$Ra colliding pairs, the presence of pairing correlations prevents the occurrence of ternary quasifission. The effect is not as striking in the cases of tail-to-tail  collisions of octupole deformed nuclei, nevertheless pairing has a pronounced influence of the location of the energy window in which the formation of a middle fragment is predicted to take place. For tail-to-tail and tail-to-side collisions, model calculations predict the formation of very heavy neutron-rich systems in certain energy intervals, a result that is potentially interesting in studies of synthesis of superheavy elements.

\begin{acknowledgments}
This work has been supported in part by the High-end Foreign Experts Plan of China, the National Natural Science Foundation of China (Grants No. 11935003, 11975031, 12070131001, and 12141501), the High-performance Computing Platform of Peking University, the QuantiXLie Centre of Excellence, a project co-financed by the Croatian Government and European Union through the European Regional Development Fund - the Competitiveness and Cohesion Operational Programme (KK.01.1.1.01.0004), and the Croatian Science Foundation under the project Uncertainty quantification within the nuclear energy density framework (IP-2018-01-5987). We acknowledge the funding support from the State Key Laboratory of Nuclear Physics and Technology, Peking University. Z. X. Ren is supported in part by the European Research Council (ERC) under the European Union's Horizon 2020 research and innovation programme (Grant agreement No. 101018170).
\end{acknowledgments}

%\bibliography{myref}

\end{document}